\documentclass[11pt,a4paper]{article}
\usepackage[margin=1in]{geometry}
\usepackage[T1]{fontenc}
\usepackage[utf8]{inputenc}
\usepackage{lmodern}
\usepackage{newunicodechar}
\newunicodechar{−}{\ensuremath{-}}

\usepackage{amsmath,amssymb,amsthm,amsfonts,mathtools}
\usepackage{bm,nccmath,mathrsfs,dsfont,bbold,slashed}
\usepackage{graphicx}
\graphicspath{{./}{tmpfig/}{figures/}}
\usepackage{array,multirow,booktabs,makecell,longtable,dcolumn}
\usepackage{rotating,overpic,textcomp,balance}
\usepackage{subcaption}
\usepackage[section]{placeins}
\usepackage{microtype}
\usepackage[hidelinks]{hyperref}

\title{Continuum Degeneracy and Systematic Limitations in X-Ray Pulsar Searches for Axion-Like Particles}
\author{%
Yen-Jhen Liu$^{1,2}$ \quad Yi Yang$^{1,2}$\\[0.8em]
\small $^{1}$Department of Physics, National Cheng Kung University, Tainan, Taiwan (ROC)\\
\small $^{2}$Institute of Physics, Academia Sinica, Taiwan (ROC)\\[0.4em]
\small \texttt{yenjhen.liu@gmail.com}
}
\date{}

\begin{document}
\maketitle
\begin{abstract}
Conversion of X-ray photons into low-mass axion-like particles (ALPs) in a pulsar magnetosphere can remove flux across an entire observing band without producing a narrow spectral feature. We test whether this smooth deficit can be distinguished from the intrinsic continuum in spectra from the Neutron Star Interior Composition Explorer (NICER) for PSR J2229$+$6114 (207.8~ks, 1--6~keV) and PSR J1849$-$0001 (90.6~ks, 2--7~keV). We fold a dipole conversion model that includes QED vacuum birefringence through the detector response and fit each spectrum with and without freedom in the absorbed power-law continuum. For $m_a\lesssim10^{-5}$~eV, holding the continua fixed gives conditional coupling bounds for both pulsars. Once the continua are refitted, each spectrum accommodates even a modeled 10\% flux deficit at its lower band edge. Neither the individual nor the joint scan yields a coupling exclusion over the tested weak-mixing range. A forecast for a J2229-like spectrum separates the effect of continuum freedom from that of spectral uncertainties. With the continuum fixed, a statistics-only forecast can reach the CERN Axion Solar Telescope (CAST) bound; retaining the adopted 1.5\% fractional uncertainty per grouped bin puts that bound beyond every field and exposure combination tested, up to 100 times the reference exposure. For these spectra, the observable signature is the curvature left after refitting the continuum. A larger conversion probability alone does not guarantee a measurable distortion.
\end{abstract}

\clearpage

\section{Introduction}
\label{sec:Intro}
Quantum chromodynamics (QCD) permits a CP-violating term in the Lagrangian. The strong CP problem is why its physical coefficient, $\bar\theta$, is so small, as implied by the absence of an observed neutron electric dipole moment \cite{CP,CP_1,CP_2}:
\begin{equation}
    \mathcal{L}_{\text{CP-vio}}=\frac{\alpha_{s}}{4 \pi} \bar\theta \operatorname{Tr}\left(G_{\mu\nu} \tilde{G}^{\mu\nu}\right)
\end{equation}
where $G_{\mu\nu}$ is the gluon field strength tensor and $\tilde{G}^{\mu\nu}$ its dual.
To resolve this, the Peccei-Quinn mechanism was proposed, giving rise to the axion - a pseudoscalar Nambu-Goldstone boson. Specific realizations, such as the archion model, suggest these particles may arise from spontaneously broken family symmetries, acting simultaneously as familons and Majorons \cite{CosCP}.
Extensions of the SM predict more general axion-like particles (ALPs) that couple to photons via 
\begin{equation}
\mathscr{L}_{a \gamma \gamma} = -\frac{1}{4} g_{a \gamma \gamma} a F_{\mu \nu} \tilde{F}^{\mu \nu} = g_{a \gamma \gamma} a \textbf{{E}} \cdot \textbf{B},
\end{equation}
where $a$ is the axion field, $g_{a\gamma\gamma}$ is the coupling constant, $F$ and $\tilde{F}$ are the electromagnetic field-strength tensor and its dual, \textbf{{E}} and \textbf{B} are the electric field and magnetic field. For an electromagnetic wave, only the photon polarization with \(\mathbf{E}\) along the background \(\mathbf{B}\) mixes with the ALP. This property enables photon-ALP conversion in strong magnetic fields, photons can therefore convert to ALPs as they leave the magnetosphere \cite{conversion_1, conversion_LSW}.

ALPs are probed by experiments targeting either non-relativistic or relativistic regimes. Non-relativistic ALPs are primary dark matter candidates, searched for in haloscopes like Axion Dark Matter eXperiment (ADMX) and Taiwan Axion Search Experiment with a Haloscope (TASEH) \cite{ALP_1,ALP_2,ALP_3,ALP_4,TASEH}  in resonant cavity, and by cosmology (BBN, the CMB) \cite{BBN, CMB, astrophy}. Moreover, their primordial distribution may be characterized by nonlinear inhomogeneities known as archioles \cite{Primordial,nonlinear}. Furthermore, the inflationary dynamics of ALP-related fields can lead to the formation of primordial black hole clusters, serving as seeds for supermassive black holes.
Relativistic ALPs are explored by helioscopes like CERN Axion Solar Telescope (CAST) \cite{CAST}, by light-shining-through-walls (LSW) experiments like Any Light Particle Search (ALPS) \cite{LSW}, and by pulsarscope studies \cite{Pulsarscope}.
Pulsars provide strong magnetic-field environments in which to investigate photon–ALP conversion. Young rotation-powered pulsars typically possess surface dipole magnetic fields of $10^{11}$–$10^{13}$ G, substantially stronger than those achievable in terrestrial experiments. Photon–ALP conversion in these environments can imprint an energy-dependent deficit on the emitted X-ray spectrum, potentially providing an observational signature of ALPs. However, the detectability of this signature depends not only on the conversion probability and the available photon statistics, but also on whether the resulting spectral modification can be distinguished from variations in the intrinsic X-ray continuum.
In particular, a smooth conversion-induced depletion may be partially absorbed by adjustments to the continuum normalization and spectral shape. Furthermore, improvements expected from stronger magnetic fields or longer exposures may be limited by systematic uncertainties that do not decrease with increasing photon statistics. These effects raise a broader question: under what conditions can X-ray pulsar observations provide competitive sensitivity to photon–ALP conversion, and which uncertainties ultimately limit that sensitivity?
In this work, we investigate these questions using observations from NASA's Neutron Star Interior Composition Explorer (NICER, 0.3-12 keV)~\cite{NICER}, focusing on the relativistic ALP regime. We analyze the X-ray spectra of PSR J2229$+$6114 and PSR J1849$-$0001 using a forward-folded conversion model, with the continuum parameters treated as nuisance parameters. We examine the distinguishability of the conversion signature through a sequence of continuum-parameter restrictions and assess the robustness of the resulting sensitivity under the tested model variations. We also use analytic scaling and forward-folded Asimov forecasts to examine how magnetic-field strength, effective statistics, continuum freedom, and the adopted systematic-error treatment affect the expected sensitivity.
The objective is therefore not limited to obtaining coupling constraints from the selected NICER observations, but also to identifying the conditions and limitations relevant to future X-ray pulsar searches for ALPs. Section~\ref{sec:Photon-ALP conversion} describes the photon–ALP conversion model and its applicable regime. Section~\ref{sec:Analysis} presents the observational data and analysis framework. Section~\ref{sec:Result} presents the coupling scans, continuum-freedom diagnostics, robustness tests, and sensitivity forecasts. Section~\ref{sec:Conclusion} summarizes the implications for future searches.

\section{Photon-ALP conversion}
\label{sec:Photon-ALP conversion}
Throughout this work, we work in natural units $\hbar=c=1$, with $m_{a}$ in eV and $g_{a\gamma\gamma}$ in GeV$^{-1}$. 
For radial propagation, let $R_{NS}$ be the neutron-star radius and $z$ the distance outward from its surface, so the radial coordinate is $r=R_{NS}+z$. The photon polarization that mixes with the ALP and the ALP itself form a two-state system. Their amplitudes evolve along the ray according to the mixing matrix $\mathcal{M}$:
\begin{equation}
-i\frac{d}{dz}\Psi(z) = \mathcal{M}\Psi(z), \quad \Psi(z) = \begin{pmatrix} A_{\text{mix}}(z) \\ a(z)\end{pmatrix}, \quad \mathcal{M} = \begin{pmatrix} \Delta_{\gamma}(z) & \Delta_{a\gamma}(z) \\ \Delta_{a\gamma}(z) & \Delta_{a}(z) \end{pmatrix}
\label{eq:SchEq}
\end{equation}
The off-diagonal term $\Delta_{a\gamma}(z)=g_{a\gamma\gamma}B_{\perp}(z)/2$ mixes the photon amplitude $A_{mix}$ with the ALP amplitude $a$. Here $B_{\perp}$ is the component of the external field transverse to the ray. The difference of the diagonal terms, $\Delta_{\gamma}-\Delta_{a}$, sets the relative phase accumulated per unit distance. For the surface-emission model used here, the initial state at $z=0$ is a photon. At photon energy $E$, the ALP mass and the plasma frequency $\omega_{pl}$ give the following positive wavenumber scales \cite{qterm1, qterm2}:

\begin{equation}
    q_{mass}(E)=\frac{m_a^2}{2E},
    \qquad
    q_{plasma}(z,E)=\frac{\omega_{pl}^2(z)}{2E}
\end{equation}
The ALP diagonal term is $-q_{mass}$ and the plasma contribution to the photon diagonal term is $-q_{plasma}$. To specify the latter, we take the magnitude of the Goldreich-Julian co-rotation charge density divided by the elementary charge $e$ as the reference number density \cite{GJchargedensity}, 

\begin{equation}
    n_{GJ} \simeq \frac{|\boldsymbol{\Omega}\!\cdot\!\mathbf{B}|}{2\pi c\,e},
    \qquad
    n_{e}=\kappa\, n_{GJ},
    \qquad
    \omega_{pl}^{2}=\frac{4\pi\alpha\,n_{e}}{m_{e}},
    \label{eq:plasma}
\end{equation}

where $\boldsymbol{\Omega}$ is the stellar angular velocity and $n_{e}$ is the effective lepton density entering the plasma frequency. We take $\kappa=n_{e}/n_{GJ}$ to be constant along the ray. We use $\kappa=1$ as the co-rotation density baseline and $10^{3}$ and $10^{4}$ as benchmarks for a pair-enriched plasma. These orders of magnitude occur in models of polar-cap pair cascades \cite{PairCascades}; they are not measurements of either target. Since $\omega^{2}_{pl}\propto\kappa$, these choices test how a larger plasma term changes the predicted spectral shape (Section~\ref{sec:profile}).

Along a radial ray at fixed magnetic geometry, a dipole has $B\propto r^{-3}$. At fixed $E$ and $\kappa$, it follows that $q_{plasma}\propto \omega^{2}_{pl}\propto n_{e}\propto |\mathbf{\Omega\cdot B}|\propto r^{-3}$. The vacuum contribution has a different radial dependence because it is quadratic in the transverse field. At the surface, the weak-field Euler-Heisenberg result for the two refractive indices is \cite{weakf1, weakf2}

\begin{equation}
  n_{\parallel} = 1 + \frac{7\alpha}{90\pi}
    \left(\frac{B_{0}\sin\theta}{B_{cr}}\right)^{2},
  \qquad
  n_{\perp} = 1 + \frac{2\alpha}{45\pi}
    \left(\frac{B_{0}\sin\theta}{B_{cr}}\right)^{2},
  \label{eq:nindices}
\end{equation}

 Here $\theta$ is the angle between the ray and the magnetic field, $B_{0}$ is the adopted surface field, and $B_{cr}=m_e^2/e$ is the critical field. The parallel index refers to the photon mode whose electric field lies in the plane containing the ray and the external field. This is the mode that mixes; both index shifts vanish for propagation along the field. Its surface QED contribution to the wavenumber mismatch is 

\begin{equation}
    q_{QED}=E\,(n_{\parallel}-1)=E\,K_{QED},
    \qquad K_{QED}=\frac{7\alpha}{90\pi}(\frac{B_{0}\sin\theta}{B_{cr}})^{2}\ge0.
    \label{eq:qqed}
\end{equation}

For an exterior multipole of order $l$, the field along the ray scales as $r^{-(l+2)}$. In terms of the distance $z$ above the surface, the field and its QED contribution are therefore

\begin{equation}
  B(z) = B_{0}\left(\frac{R_{NS}}{R_{NS}+z}\right)^{l+2}, \quad q_{QED}(z,E)=E\,K_{QED}(\frac{R_{NS}}{R_{NS}+z})^{2(l+2)}
  \label{eq:bprofile}
\end{equation}

For a dipole $(l=1)$, the QED term therefore decreases as $r^{-6}$, much more rapidly than the plasma contribution. The total phase mismatch used in the integral is

\begin{equation}
    q_{tot}=q_{mass}-q_{plasma}+q_{QED}.
\end{equation}

Equation~\eqref{eq:SchEq} gives a particularly simple amplitude in weak mixing. Start with $A_{mix}(0)=1$ and $a(0)=0$. To first order in $\Delta_{a\gamma}$ the photon propagates with its diagonal phase alone. Removing the ALP diagonal phase by writing $\tilde{a}(z)=a(z)\exp[-i\int^{z}_{0}\Delta_{a}(z')dz']$ gives 
\begin{equation}
    \frac{d\tilde{a}}{dz}=i\Delta_{a\gamma}(z)\exp[i\int^{z}_{0}(\Delta_{\gamma}(z')-\Delta_{a}(z'))dz']
    \label{eq:preweakmixing}
\end{equation}
Integrating the amplitude and taking its squared modulus gives the leading conversion probability:
\begin{equation}
    \mathcal{P}(\gamma\to a)=\left|\int_0^{\infty}\!dz\,\Delta_{a\gamma}(z)\,
\exp\left(i\int_0^{z} q_{\mathrm{tot}}(z')\,dz'\right)
\right|^2
\label{eq:weakmixing}
\end{equation}

\subsection{Template and applicable regime}
\label{sec:template}
To derive the energy dependence of the conversion probability in closed form, we specialize to a dipole and keep the QED term in the phase over the region that contributes to conversion. Write $R=R_{NS}$ and introduce the dimensionless radius $x=r/R=1+z/R$. This equality follows from $r=R+z$. At fixed propagation angle, $B_{\perp,0}=B_{0}\sin\theta$ is the transverse surface field. Setting $l=1$ in Equation \eqref{eq:bprofile} gives $B_{\perp}\propto x^{-3}$; the QED mismatch is quadratic in this field and the off-diagonal mixing is linear in it. Thus

\begin{equation}
B_{\perp}(x) = B_{\perp,0}x^{-3}, \quad q_{QED}(x) = K_{QED}E_{loc}x^{-6}, \quad \Delta_{a\gamma}(x) = \frac{g_{a\gamma\gamma}B_{\perp,0}}{2}x^{-3}
\label{eq:dipole-field}
\end{equation}
Here $K_{QED}$ is the surface coefficient defined above, and 
$E_{loc}$ is taken to be constant along the trajectory in this approximation. With $q_{tot}\simeq q_{QED}$, the accumulated phase is
\begin{equation}
    \int^{z}_{0}q_{tot}(z')dz'\simeq\frac{K_{QED}E_{loc}R}{5}(1-x^{-5})\equiv A(1-x^{-5}).
\end{equation}
Substitution into Eq.~\eqref{eq:weakmixing}, followed by the change of variable $t=x^{-5}$, gives 
\begin{equation}
    P_{\gamma\to a} \simeq g^{2}_{a\gamma\gamma}(\frac{B_{\perp,0}R}{2})^{2}|\frac{e^{iA}}{5}\int^{1}_{0}t^{-3/5}e^{-iAt}dt|^{2}.
\end{equation}
For $A\gg1$, substituting $u=At$ gives the leading term by extending the upper limit in $u$ to infinity. In terms of the Euler gamma function $\Gamma(s)$, this is 
\begin{equation}
    \int^{1}_{0}t^{-3/5}e^{-iAt}dt\simeq\Gamma(\frac{2}{5})e^{-i\pi/5}A^{-2/5}.
    \label{eq:intt}
\end{equation}
Taking the squared modulus gives the conversion probability. To make its use in the spectral fit explicit, we factor out the coupling and denote the remaining energy-dependent coefficient by $\hat{P}(E)$:
\begin{equation}
    \mathcal{P}_{\gamma\to a} (E) \simeq g_{a\gamma\gamma}^{2}\underbrace{\left(\frac{B_{\perp,0}R_{NS}}{2}\right)^{2}
    \left[\frac{\Gamma(2/5)}{5}\right]^{2}\left(\frac{K_{QED}\,E_{loc}\,R_{NS}}{5}\right)^{-4/5}}_{\hat{P}(E)}.
    \label{eq:phat}
\end{equation}

At this order the amplitude is linear in $g_{a\gamma\gamma}$, and the phase is independent of it. Varying the coupling therefore changes the amplitude of the predicted depletion without changing its spectral shape.
To express the template in terms of the observed energy, we use the constant surface-lapse approximation
\begin{equation}
E_{loc} = \frac{E}{\alpha_{g}}, \quad E \equiv E_{obs} , \quad \alpha_{g}=\sqrt{1-\frac{2GM_{NS}}{R_{NS}}}.
\label{eq:surface-lapse}
\end{equation}
Thus the lapse contributes a factor $\alpha^{4/5}_{g}$ to the template normalization, while its energy dependence remains $\hat{P}(E)\propto E^{-4/5}$. This is an approximate treatment of gravitational effects: the local photon energy and proper distance can vary along the trajectory in a more complete calculation.

The result is a large-phase, far-zone approximation rather than a general consequence of a dipolar field. Its characteristic radius is $r_{*}\sim RA^{1/5}$, where the remaining QED phase becomes of order unity. Its use requires $A\gg1$, a dipolar field extending well beyond this radius, and mass and plasma terms that do not appreciably change the relative phase over the region contributing to the integral. In particular, QED dominance at the stellar surface alone does not establish these conditions: the QED term falls as $r^{-6}$, faster than the plasma term, while the mass term does not fall with radius in the constant-energy approximation. If the light cylinder, plasma, or mass term affects the contributing region, the full propagation integral or an appropriate numerical calculation should be used instead. The weak-mixing treatment must also be self-consistent, with the resulting conversion probability much smaller than unity.

\subsection{Polarization}
\label{sec:polarization}
 Since only the ordinary mode (O-mode) couples to the ALP, the magnitude of the flux deficit depends on how much of the emitted light occupies that mode. We denote its fraction by $f_{O}$. The extraordinary mode (X-mode) propagates without mixing and dilutes the deficit in the total spectrum. Under the two-mode polarization description adopted here, the linear polarization degree is ${\rm PD}=|2f_{O}-1|$, so $f_{O}=(1\pm{\rm PD})/2$, and unpolarized emission corresponds to $f_{O}=1/2$. These relations are used to express the polarization dependence of the forward model, not as measurements of either target. The baseline coupling scans are performed with $f_O=1$. We assume $f_O$ to be energy-independent across the fitted band; under this assumption the model depends on $f_O$ and $g_{a\gamma\gamma}$ only through $f_Og_{a\gamma\gamma}^{2}$, so results for other $f_O>0$ follow from the exact scaling $g_{a\gamma\gamma}\propto f_{O}^{-1/2}$. More generally, any unmodeled non-converting flux component with a similar spectral shape, such as unresolved pulsar wind nebula (PWN) emission in phase-averaged NICER spectra, acts equivalently to reducing the effective converting fraction $f_O$, and its dilution effect is captured by the same $f^{-1/2}_{O}$ scaling. We use $f_O=1/2$ only when quoting the conditional unpolarized reference values, as stated in Section~\ref{sec:profile-result}.

\section{Data and analysis framework}
\label{sec:Analysis}

\subsection{Observations, fitted bands and source continua}
\label{sec:obs}
We analyze NICER observations of the rotation-powered pulsars PSR J2229$+$6114 and PSR J1849$-$0001. All data were reduced with HEASoft 6.35.2 and NICERDAS v.11, beginning with \texttt{nicerl2} under the standard screening criteria, followed by spectral extraction with \texttt{nicerl3-spect}. Backgrounds were estimated with the empirical 3C50 model~\cite{3c50}. For each source we use the ten selected observations listed in Appendix~\ref{app:data_table}, with total cleaned exposures of $207.8$ks for PSR~J2229$+$6114 and $90.6$ks for PSR~J1849$-$0001, respectively. Table~\ref{tab:obs} summarizes the source parameters and fitted energy bands.

\begin{table}[tbp]
\centering
\begin{tabular}{lcccc}
\toprule
Pulsar & $B_{0}$ [G] & Spin period $P$ [s] & Band [keV] & Exposure [ks] \\
\midrule
PSR~J2229$+$6114 & $2.03\times10^{12}$ & 0.0516 & 1.0-6.0 
  & $207.8$ \\
PSR J1849$-$0001 & $7.5 \times10^{11}$ & 0.0385
  & 2.0-7.0 
  & $90.6$ \\
\bottomrule
\end{tabular}
\caption{NICER observations and magnetospheric parameters. The values of $B_{0}$ are fiducial spin-down-inferred dipole field estimates~\cite{pulsarB,J2229,J1849}, adopted as the surface-field normalizations and held fixed in the baseline analysis. They are model-dependent estimates rather than direct measurements of the local magnetic field. The spin period is $P$; the band is the energy range used in the spectral fit.}
\label{tab:obs}
\end{table}

Each spectrum is described by an absorbed power law, \texttt{TBabs}$\times$\texttt{powerlaw}, with abundances following Wilms~et~al.~\cite{wilms2000}. The free continuum parameters are the hydrogen column density $n_{H}$, photon index $\Gamma$, and normalization $N$. The continuum is fitted globally across each source's band. The fitted bands are $1-6$ keV for J2229$+$6114  and $2-7$ keV for J1849$-$0001. For the more strongly absorbed J1849$-$0001, the simple absorbed power law does not adequately describe the lower-energy turnover. 
A third source, PSR~B0531$+$21 (the Crab), was considered but is not included in the reported search because the simple continuum adopted here does not provide an adequate description of its spectrum over the intended fit range. A conversion search requires both a specified conversion template and a continuum model appropriate to the data.

\subsection{Predicted NICER counts and spectral fitting}
\label{sec:countspace}
At each trial coupling we calculate how photon conversion changes the flux incident on NICER. The effective area sets the number of photons collected, and the energy redistribution assigns them to measured channels. Let $d$ label a spectrum with its associated exposure and instrument response, and $i$ a channel in measured pulse-invariant energy (or a group of channels used in the fit). The predicted counts before background subtraction are

\begin{equation}
\begin{split}
  \mu_{di} = t_{d}\int\!\mathrm{d}E\;R_{di}(E)\,A_{d}(E)\,
  \Big\{\; &F^{0}_{X}(E;\boldsymbol{\eta}) \\[2pt]
  &+ \big[1-\mathcal{P}_{\gamma\to a}(E;m_{a},g_{a\gamma\gamma},\boldsymbol{\xi})\big]\,
     F^{0}_{O}(E;\boldsymbol{\eta}) \Big\}
  + b_{di}(\boldsymbol{\beta}),
\end{split}
\label{eq:forward}
\end{equation}

where $t_{d}$ is the exposure, $A_{d}(E)$ is the effective area, and $R_{di}(E)$ is the probability for a detected photon of true energy $E$ to be recorded in channel $i$ of spectrum $d$. The quantities $F^{0}_{X}$ and $F^{0}_{O}$ are the incident fluxes in the two modes before conversion, $\boldsymbol{\eta}$ denotes the continuum parameters, $\boldsymbol{\xi}$ denotes the magnetospheric parameters, and $b_{di}(\beta)$ is the modeled background count contribution.
We assume the two normal modes share a single spectral shape, so that $F^{0}_{X}=(1-f_{O})F^{0}$ and $F^{0}_{O}=f_{O}F^{0}$, with $F^{0}$ the absorbed power law. Equivalently, $f_{O}$ is independent of energy across the fitted band. Equation~\eqref{eq:forward} then reduces to 

\begin{equation}
  \mu_{di} = t_{d}\int\!\mathrm{d}E\;R_{di}(E)\,A_{d}(E)\,
  F^{0}(E;\boldsymbol{\eta})
  \Big[1 - f_{O}\,g_{a\gamma\gamma}^{2}\,\hat{P}(E;m_{a},\boldsymbol{\xi})\Big]
  + b_{di}(\boldsymbol{\beta}),
  \label{eq:forward-reduced}
\end{equation}

We fit the folded source prediction to the grouped, 3C50-subtracted NICER counts, using the default optimal binning of \texttt{nicerl3-spect}. The continuum parameters are shared across the fitted band of each source. The $\chi^{2}$ fit uses the statistical and supplied systematic uncertainties in the spectral products without further rescaling. The background term in Equation~\eqref{eq:forward} specifies the contribution before subtraction; uncertainty in the 3C50 background prediction is not profiled separately.

\subsection{Conversion template: implementation and range of validity}
\label{sec:numericalVal}
Equation~\eqref{eq:phat} specifies the conversion template $\hat{P}(E)$. We implement the corresponding survival factor in Equation~\eqref{eq:forward-reduced} as a multiplicative \texttt{mdefine} component on \texttt{TBabs}$\times$\texttt{powerlaw} in XSPEC. We validate the conversion calculation independently with a fourth-order Runge-Kutta (RK4) propagation of the photon-ALP system. The comparison is performed at four energies across each fitted band and for $\kappa=1, 10^{3}$ and $10^{4}$. Details of the numerical propagation and the resulting comparison are given in Appendix~\ref{app:B}. 
For the pair-multiplicity study, where the plasma term modifies the spectral shape, we evaluate the phase integral numerically and approximate the resulting spectrum in a form suitable for XSPEC. A $\kappa$-dependent diagnostic is retained only when this representation reproduces the numerical template to better than $2\%$ across the fitted band. 
The representation accuracy above concerns the numerical implementation. A separate restriction limits the physical use of the leading-order probability in Equation~\eqref{eq:forward-reduced}. We restrict the baseline scan to the coupling at which the predicted depletion reaches $10\%$ at the low-energy edge of each fitted band, where the baseline $E^{-0.8}$  template is largest. For $m_{a}=10^{-6}$eV, $\theta=90^\circ$, and the baseline $f_O=1$ convention, this gives $\log_{10}(g_{a\gamma\gamma}/GeV^{-1})\leq-9.549$ for J2229$+$6114, and $\leq-9.342$ for J1849$-$0001. We additionally repeat the continuum-freedom diagnostic using $5\%$ and $15\%$ caps to test the dependence on this choice.

\subsection{Coupling scans, simulations, and robustness tests}
\label{sec:profile}
For each trial value of the coupling we refit the continuum parameters $\boldsymbol{\eta}=(n_{\rm H}, \Gamma,N)$ and compare the fit quality with that at the near-zero reference coupling. We scan 41 logarithmically spaced points from $10^{-11.5}\,\mathrm{GeV}^{-1}$ up to the depletion cap for each source. We measure the change in fit quality by 
\begin{equation}
    \Delta\chi_{0}^{2}(g_{a\gamma\gamma})=\min_{\eta}\chi^2(g_{a\gamma\gamma},\eta)-\min_{\eta}\chi^2(g_{\rm ref,0},\eta),\qquad g_{\rm ref,0}=10^{-11.5}\ \mathrm{GeV}^{-1}.
    \label{fig:minchi}
\end{equation}
A positive value means the fit is worse than the reference, and a negative value means it is better. The subscript 0 denotes the near-zero-coupling reference at $g_{\rm ref,0}=10^{-11.5}\mathrm{GeV}^{-1}$, rather than necessarily the best fit over the scanned coupling range. We adopt $\Delta\chi^{2}_{0}=2.71$ as a threshold for the scan. This value is nominally associated with a one-sided $95\%$ criterion under standard statistical assumptions, but we do not claim calibrated $95\%$ coverage for the present procedure. We report whether the scan reaches this threshold within the perturbatively valid range. The two sources are treated as independent. Their joint statistic is obtained by summing the individual profiled differences over the common valid coupling domain, with interpolation onto a shared grid where required.

To measure how often such a scan yields a threshold crossing when conversion is negligible, we simulate 2,000 spectra for each pulsar at the near-zero reference coupling using XSPEC \texttt{fakeit}. The simulations use the best-fit continuum, instrumental response, background treatment, exposure, and the same free-continuum scan as the observed spectra. The generated source counts include Poisson fluctuations, and the simulated background is scattered according to the statistical uncertainties supplied in the background product. In the primary ensemble, the NICER systematic-error vector is retained in the fit uncertainty but is not used to add an extra random perturbation to the generated counts. A separate stress-test ensemble adds Gaussian fluctuations at the supplied systematic-error level. These simulations do not quantify an unknown bias in the 3C50 background prediction or an unknown calibration offset. The diagnostic is the fraction of realizations whose scan reaches the adopted threshold within the valid coupling range; it is not a confidence-interval coverage study at injected nonzero coupling.

Because the continuum parameters are correlated, changes between successive rungs diagnose the effect of the specified sequence and should not be interpreted as unique orthogonal contributions of the individual parameters.
To determine how much of the result depends on nuisance-parameter freedom, we repeat the scan with nested subsets of the continuum held fixed at their no-signal best-fit values: (i) all three free; (ii) $n_{\rm H}$ fixed; (iii) $n_{\rm H}$ and $\Gamma$ fixed; and (iv) all three fixed. The data, the conversion template and the fit band and scanned coupling range are identical in all four treatments. 
We compare the maximum statistic difference over that range. Crossings obtained with fixed continuum parameters are conditional on those parameters being known and do not incorporate their uncertainties. No independent determination of the continuum parameters is supplied for these data.

Three classes of variation are considered. First, parameters that only rescale the fixed low-mass template obey exact or analytic coupling rescaling: 
$ g_{a\gamma\gamma}\propto f_{O}^{-1/2}, g_{a\gamma\gamma}\propto \sin^{-1/5}\theta, g_{a\gamma\gamma}\propto B_{0}^{-0.2}$
at fixed stellar radius. 
Second, pair multiplicity is  treated with explicit plasma-aware templates because it can change the template shape.
Third, the observed-data continuum-freedom ladder is repeated with $5\%$, $10\%$, and $15\%$ depletion caps. 
The adopted 3C50 background model is not varied in these tests.

\subsection{Forecast framework}
\label{sec:forecast}
The conditional fixed-continuum crossing provides an anchor for an idealized competitiveness forecast. We define a dimensionless effective-statistics multiplier
\begin{equation}
    n \equiv N_{eff}/N_{eff,ref}
\end{equation}
where $n=1$ denotes the J2229$+$6114 reference and $n$ represents a relative effective statistical multiplier for the fixed template. In the idealized scaling considered here, an increase in $n$ may be interpreted as an exposure-equivalent increase for a J2229-like spectrum with the spectral shape, response, and background fraction held fixed. 
For a J2229-like spectrum in the low-mass regime, the ALP-induced count change scales as $\delta\mu_{i}\propto nf_{O}g_{a\gamma\gamma}^{2}B_{0}^{0.4}$ at $\theta=90^{\circ}$. In the statistic-dominated limit the variance scales as $\sigma_{i}^2\propto n$. Therefore
\begin{equation}
    \Delta\chi^{2}\propto(\delta\mu)^{2}/\sigma^{2} \propto f_{O}^{2}g_{a\gamma\gamma}^{4}B_{0}^{0.8}n.
\end{equation}
At a fixed threshold, this gives the ideal sensitivity scaling
\begin{equation}
    g_{\rm thr} = g_{ref} f_{O}^{-1/2}(B_{0}/B_{ref})^{-1/5}n^{-1/4}
    \label{eq:g95}
\end{equation}

For the 1–6 keV J2229$+$6114 analysis, the observed-data, all-fixed threshold crossing is $g_{\rm ref}=1.4139\times10^{-10}\mathrm{GeV}^{-1}$ at $f_{O}=1$ and $\theta=90^{\circ}$, with $B_{\rm ref}=2.03\times10^{12}\mathrm{G}$. We define
\begin{equation}
    R_{CAST}\equiv g_{CAST}/g_{\rm thr},
\end{equation}
Values of $R_{CAST}>1$ correspond to a threshold-crossing reference stronger than the CAST comparison bound. This  scaling assumes a fixed continuum, statistics-dominated uncertainties, and the weak-field QED. The region above $B_{cr}\simeq4.4\times10^{13}$G is an extrapolation beyond the weak-field treatment. 
To evaluate the effects of  continuum freedom and the adopted systematic error treatment, we construct a forward-folded Asimov spectrum from the best-fit J2229$+$6114 continuum, response, and background model. The spectrum represents the expected counts without random fluctuations. We scale the source and background counts together with exposure while retaining the spectral shape, response, and background fraction. In rate space, the source and background statistical variances scale as $1/n$. For the systematic error treatment, the supplied fraction \texttt{SYS\_ERR} is applied to the source-plus-background counts and remains fixed as exposure increases; the fit uses a diagonal variance.
We evaluate 673 points, including the reference point, over $B_{0}=10^{12}$ to $3.76\times10^{13}\mathrm{G}$ and $n=0.3$ to 100. The core region extends to $2\times10^{13}\mathrm{G}$; higher-field points are treated as a caution region. At each point, we apply the four continuum treatments described in Section~\ref{sec:profile} and scan 41 logarithmically spaced coupling values up to the corresponding depletion cap. Threshold crossings are interpolated in $\log_{10}g_{a\gamma\gamma}$. The magnetic field dependence is implemented through the template-amplitude scaling $\hat{P}\propto B_{0}^{0.4}$.
 
\section{Results}
\label{sec:Result}

\subsection{Baseline fits and coupling scans}
\label{sec:baseline}
\begin{figure}[tbp]
\centering
\includegraphics[width=\textwidth]{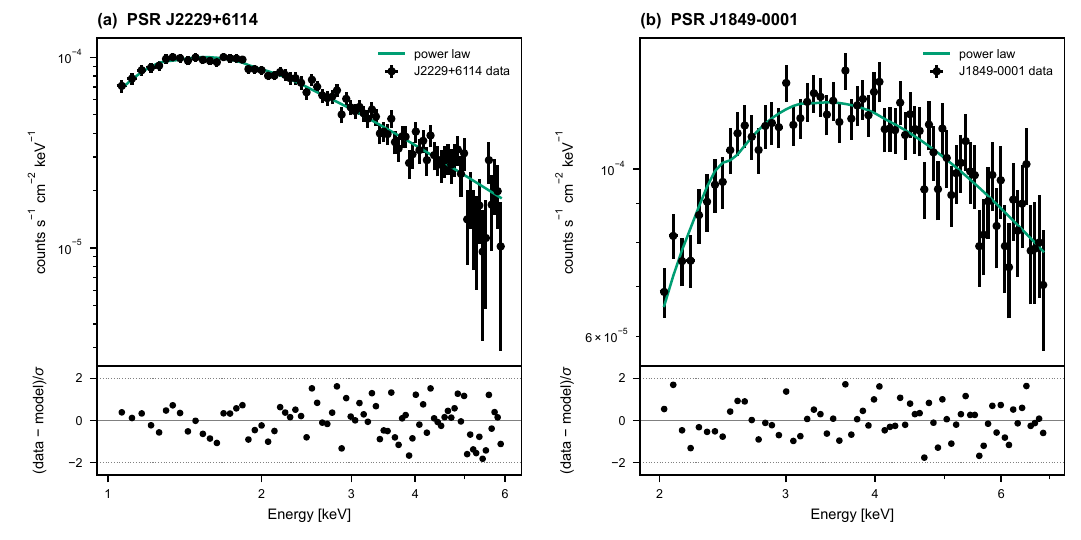}
\caption{NICER spectra of PSR J2229$+$6114  (left, 1–6 keV) and PSR J1849$-$0001 (right, 2–7 keV). Each upper panel shows the background-subtracted, area-corrected count-rate spectrum and its absorbed-power-law model. Below each spectrum, one point per grouped energy bin shows the pull, $(\rm{data-model})/\sigma_{i}$, with the same energy scale; $\sigma_{i}$ includes the supplied systematic error. The horizontal line denotes zero. Fits are performed in count space.}
\label{fig:continuum}
\end{figure}

The absorbed power-law fits give $\chi^{2}$/dof = 54.06/77 for J2229$+$6114 and 45.20/65 for J1849$-$0001. Table~\ref{tab:continuum} lists the best-fit parameters and Figure~\ref{fig:continuum} shows the folded spectra and residuals. 

\begin{table}[tbp]
\centering
\begin{tabular}{lcccc}
\toprule
Pulsar & $n_{\rm H}$ [$10^{22}\,$cm$^{-2}$] & $\Gamma$ &
  $N$ [$10^{-4}$] & $\chi^{2}/\mathrm{dof}$ \\
\midrule
PSR J2229$+$6114 & $1.123^{+0.063}_{-0.062}$ & $1.735^{+0.056}_{-0.055}$ & $4.031^{+0.266}_{-0.247}$
  & $54.06/77$ \\
PSR J1849$-$0001 & $5.901^{+0.393}_{-0.387}$ & $1.168^{+0.084}_{-0.083}$ & $7.811^{+1.151}_{-0.995}$
  & $45.20/65$ \\
\bottomrule
\end{tabular}
\caption{Best-fit \texttt{TBabs}$\times$\texttt{powerlaw} continua in count space using the grouped, 3C50-subtracted spectra. $N$ is the power-law normalization in
photons\,keV$^{-1}$\,cm$^{-2}$\,s$^{-1}$ at 1\,keV. $1\sigma$ uncertainties from $\Delta\chi^{2}=1$,  using the adopted fit uncertainties.}
\label{tab:continuum}
\end{table}

\begin{figure}[tbp]
\centering
\includegraphics[width=0.9\textwidth]{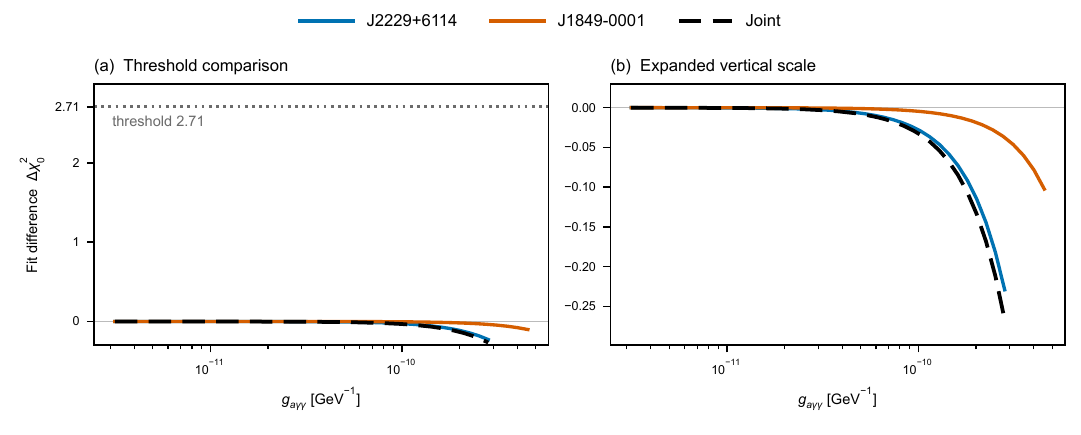}
\caption{Change in fit quality relative to the near-zero reference after refitting the continuum at each coupling, as defined in Equation~\eqref{fig:minchi}. The left panel shows the full comparison to the adopted threshold (2.71); the right panel magnifies the same curves around zero. The black dashed curve shows the joint scan. The individual curves stop at their $10\%$ depletion caps and the joint curve at the end of their common range. Negative values indicate a better fit than at the reference; the endpoints are not interior minima.}
\label{fig:profile}
\end{figure}

Figure~\ref{fig:profile} shows the free-continuum coupling scans and their joint statistic. Neither individual source nor the joint scan reaches the adopted threshold within the perturbatively valid domain. The profiles therefore provide no coupling exclusion under the adopted procedure.
For both pulsars, the profiled statistic decreases across the sampled coupling range and reaches $\Delta\chi^{2}_{0}=-0.229$ for PSR J2229+6114 and $-0.102$ for PSR J1849−0001 at their respective depletion caps. The joint statistic reaches $\Delta\chi^{2}_{0}=-0.267$ at the upper end of the common coupling range. Because the individual profiles are still decreasing at their respective caps, the scans do not identify an interior minimum. These endpoints should not be interpreted as localized preferred couplings, and we do not quote best-fit couplings from these scans.
In the 2000 near-zero-coupling simulations described in Section~\ref{sec:profile}, none of the 2000 spectra per source crossed the threshold in the primary ensemble. The separate systematic-scatter stress test also gave zero crossings for each pulsar. For 0 crossings in 2000 independent trials, the one-sided $95\%$ binomial upper bound on the crossing probability is approximately $0.15\%$. This bounds the frequency of crossing under the simulation assumptions; it  does not establish confidence-interval coverage at nonzero coupling.

\subsection{Continuum degeneracy, robustness, and conditional sensitivity}
\label{sec:profile-result} 

The absence of a threshold crossing in a coupling scan does not, by itself, establish whether the conversion signature is intrinsically too weak to be resolved or whether its spectral modification becomes indistinguishable from variations in the underlying continuum. Distinguishing these possibilities is essential for interpreting the observational null result and evaluating the prospects of future searches. We therefore apply the continuum-freedom ladder defined in Section~\ref{sec:profile}, using the same data and conversion template while progressively restricting the continuum parameters.

Table~\ref{tab:ladder} summarizes the maximum statistic differences obtained under the four continuum treatments. With all continuum parameters fixed, the maximum $\Delta\chi^{2}$ reaches $43.33$ for PSR J2229$+$6114 and $49.80$ for PSR J1849$-$0001, demonstrating substantial conditional sensitivity to the imposed conversion template. In contrast, when the continuum parameters are fully profiled, neither source reaches the adopted threshold within the investigated coupling range.
This contrast shows that the limited sensitivity of the profiled searches cannot be attributed simply to insufficient photon statistics. Instead, the conversion-induced modification can be largely accommodated by changes in the fitted continuum. For a smooth low-mass conversion template, the modified spectrum $F(E)[1-CE^{-0.8}]$ can be approximated over a finite energy band by adjustments to the amplitude and slope of the underlying power-law spectrum $NE^{-\Gamma}$. Although this degeneracy is not exact, the remaining spectral differences are insufficient to reach the adopted threshold in the fully profiled scans.
 Along the specified sequence of continuum restrictions, fixing $n_{H}$ alone leaves the maximum statistic difference at zero for both pulsars. When both $n_{H}$ and $\Gamma$ are fixed while the normalization remains free, the maximum statistic differences reach 3.170 and 2.880 for PSR J2229$+$6114 and PSR J1849$−$0001, respectively, exceeding the adopted threshold of 2.71 in both cases. The corresponding threshold crossings are $2.709\times10^{-10}$ and $4.469\times10^{-10}\mathrm{GeV}^{-1}$, respectively. Fixing the normalization as well produces the largest further increase in the maximum statistic difference. Because the continuum parameters are correlated, these changes should be interpreted as the effects of the specified nested restrictions rather than independent contributions from individual parameters. The result illustrates a central difficulty for searches based on smooth spectral depletion: substantial sensitivity under an assumed known continuum does not necessarily translate into meaningful sensitivity when the continuum must be determined from the same observations. Consequently, increasing photon statistics alone cannot be assumed to restore sensitivity without considering the distinguishability of the conversion template from the continuum model.

\begin{table}[tbp]
\centering
\begin{tabular}{lcc}
\toprule Continuum treatment & J2229$+$6114 & J1849$-$0001 \\  
\midrule All free ($n_{\rm H},\Gamma,N$) & 0.000 & 0.000 \\ $n_{\rm H}$ fixed               & 0.000 & 0.000 \\ $n_{\rm H},\Gamma$ fixed        & 3.170 & 2.880 \\ All fixed                       & 43.332 & 49.804\\
\bottomrule
\end{tabular}
\caption{Reported maximum $\Delta\chi^{2}_{0}$ over the same capped coupling range for the four nested continuum treatments. Values of 0.000 occur when $\Delta\chi^2\leq0$ across the entire scanned range, with the maximum attained at the near-zero reference coupling $g_{ref,0}$}
\label{tab:ladder}
\end{table}

\begin{figure}[tbp]
\centering
\includegraphics[width=\textwidth]{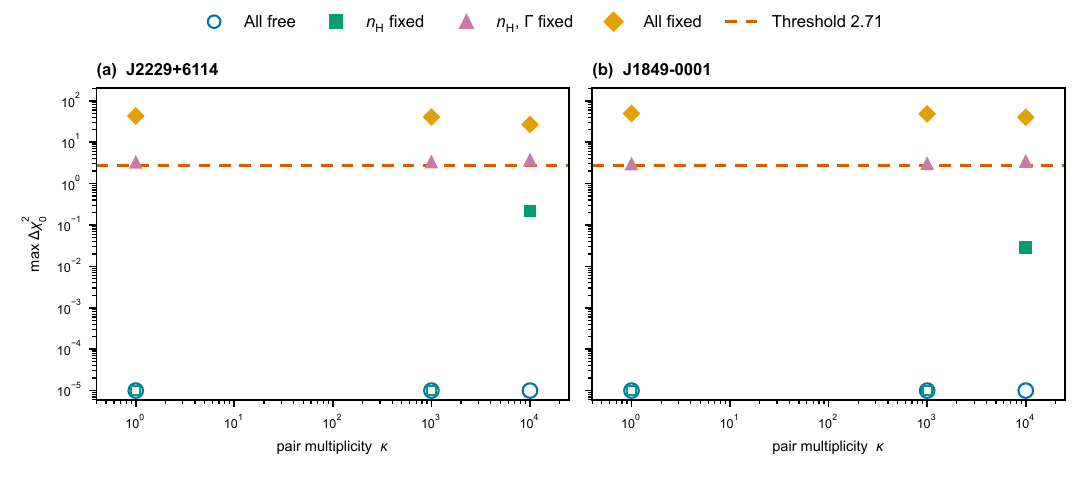}
\caption{Maximum $\Delta\chi_{0}^{2}$ from the continuum-freedom ladder for $\kappa=1$, $10^{3}$, and $10^{4}$,  for PSR J2229+6114 in the 1–6 keV band and PSR J1849$-$0001 in the 2–7 keV band. The dashed line marks the adopted threshold $\Delta\chi_{0}^{2}=2.71$. Open markers at $10^{-5}$ indicate values below the plotting floor on the logarithmic axis. The points correspond to separately evaluated multiplicities and are not connected by lines.}
\label{fig:kappa}
\end{figure}

Repeating the continuum-freedom ladder with $5\%$, $10\%$, and $15\%$ depletion caps does not change the free-continuum conclusion for either source. The all-fixed crossing couplings are likewise stable at approximately $1.41\times10^{-10}$ and $2.19\times10^{-10}\mathrm{GeV}^{-1}$ in the baseline $f_{O}=1$ convention; only the maximum statistic reached at the endpoint changes with the cap.
Figure~\ref{fig:kappa} shows the ladder at $\kappa=1$, $10^{3}$ and $10^{4}$ for both pulsars. All final XSPEC representations satisfy the $2\%$ numerical criterion; the largest verified representation error is $0.18\%$. At every tested multiplicity, the free-continuum scan remains below the adopted threshold.
The evidence therefore supports the profiled null result over the sampled pair multiplicities. It does not establish the same behavior for arbitrary plasma profiles or resonant regimes outside the tested cases. The uncertainty in the 3C50 background prediction was not independently quantified, so these results remain conditional on the adopted background model.

\begin{figure}[tbp]
\centering
\includegraphics[width=0.65\textwidth]{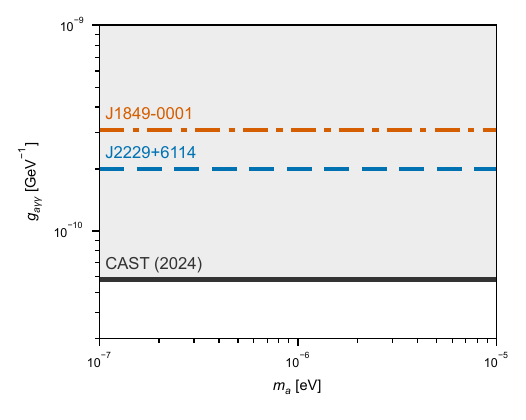}
\caption{Fixed-continuum conditional threshold-crossing references for unpolarized emission, $f_O=1/2$, compared with the CAST bound. The directly labeled NICER lines use the low-mass template and are approximate references over the displayed mass range; the residual mass dependence is discussed in Appendix \ref{app:B}.}
\label{fig:panel}
\end{figure}

With the continuum fully fixed, the baseline $f_{O}=1$ scans cross the threshold at $1.41\times10^{-10}\mathrm{GeV}^{-1}$ for J2229$+$6114 and $2.19\times10^{-10}$ GeV$^{-1}$ for J1849$-$0001. For unpolarized emission, $f_{O}=1/2$, the corresponding values are approximately $2.0\times10^{-10}\mathrm{GeV}^{-1}$ and $3.1\times10^{-10}$ GeV$^{-1}$, obtained using the scaling in Section~\ref{sec:polarization}.
Figure~\ref{fig:panel} compares these low-mass references with the CAST bound of $5.8\times10^{-11}\mathrm{GeV}^{-1}$ ~\cite{CAST}. The residual mass dependence is discussed in Appendix~\ref{app:B}, and the dependence on $f_{O}$ is shown in Appendix~\ref{app:C}.

\subsection{Can a J2229-like spectrum reach CAST sensitivity?}
\label{sec:expected}

At the J2229$+$6114 reference field and effective statistics, $R_{CAST}\simeq 0.41$ for $f_{O}=1$ and $R_{CAST}\simeq0.29$ for $f_{O}=1/2$. Equation~\eqref{eq:g95} implies the CAST boundary
\[
 n_{CAST}(B_{0},f_{O}) = \left(\frac{g_{ref}}{g_{CAST}}\right)^{4}
 f_{O}^{-2}\left(\frac{B_{0}}{B_{ref}}\right)^{-0.8}.
\]
At the current J2229 field, this corresponds to $n\approx35.3$ for $f_{O}=1$ and $n\approx141.3$ for unpolarized emission. At fixed statistics, the same ideal scaling would require approximately 86 times the current J2229 field for $f_{O}=1$ and approximately 487 times for $f_{O}=1/2$, both above $B_{cr}$.
As an approximate comparison, applying the same scaling to the J1849$-$0001 field and the ratio of the exposures listed in the Table~\ref{tab:obs} gives a conditional coupling of approximately $2.12\times10^{-10}\mathrm{GeV}^{-1}$, compared with the observed data all-fixed crossing of $2.19\times10^{-10}\mathrm{GeV}^{-1}$. 
The sources differ in spectral shape, response, background fraction, and fitted band, so this exposure-based comparison is approximate.

We evaluate a 673-point Asimov grid for a J2229-like spectrum. In the statistics-only, fixed-continuum case, the forecast reproduces the scaling of Equation~\eqref{eq:g95}. Over the core magnetic-field region, the fitted logarithmic slopes are -0.199 in $B_{0}$ and -0.249 in $n$, compared with the analytic expectations $-0.2$ and $-0.25$. Normalizing the analytic scaling to the statistics-only Asimov reference, the median fractional deviation is $0.30\%$ and the 90th-percentile deviation is $0.66\%$. At $n=1$, the statistics-only Asimov threshold crossing is $1.229\times10^{-10}\mathrm{GeV}^{-1}$, approximately $13\%$ below the observed-data all-fixed crossing of $1.414\times10^{-10}\mathrm{GeV}^{-1}$. The two references are evaluated with different uncertainty treatments: the observed-data fit includes the supplied systematic errors, whereas the  statistics-only Asimov calculation does not. For the observed spectrum, the forward-folded implementation reproduces the XSPEC null $\chi^2$ and the continuum-freedom profile curves; the largest difference in $\Delta\chi^2$ across the four treatments is $2.4\times10^{-3}$. The threshold crossings agree within $0.02\%$. No fit failures occur in either forecast grid.

With the adopted per-bin \texttt{SYS\_ERR} treatment, the sensitivity gain with increasing $n$ becomes substantially weaker. Over the core magnetic-field region, the fitted $n$ slope of the all-fixed sensitivity is approximately $-0.07$, compared with $-0.25$ in the statistics-only case. At fields near that of J2229+6114, $R_{\rm CAST}$ approaches approximately 0.5 as $n$ increases. Across the full $B_0$--$n$ grid, no point reaches the CAST comparison bound under the adopted systematic-error treatment. The maximum $R_{\rm CAST}$ is approximately 0.82 in the core magnetic-field region and 0.93 when the higher-field caution region is included.
In the 1--6 keV band, the adopted fractional systematic uncertainty is 1.5\% per grouped bin, while modeled background accounts for approximately 68\% of the total in-band counts. For illustration, a bin with this band-averaged background fraction would have a systematic uncertainty on the total counts equivalent to approximately 4.7\% of its source counts. The actual background fraction varies across the band, and the forecast evaluates the uncertainty in each bin. As the exposure increases, the statistical contribution decreases, whereas the adopted fractional systematic contribution remains fixed. The resulting sensitivity gain therefore approaches saturation.
Continuum profiling imposes an additional limitation. With $n_H$, $\Gamma$, and $N$ all free, the adopted threshold of $\Delta\chi^2_0=2.71$ is not reached anywhere on the grid under either uncertainty treatment, including at $n=100$. The largest statistic difference is approximately 0.73. Fixing $n_H$ alone produces no CAST-competitive point. When $n_H$ and $\Gamma$ are fixed and the normalization remains free, approximately $5\%$ of evaluated forecast points reach or surpass the CAST comparison bound in the statistics-only treatment; none does so under the adopted systematic-error treatment. These competitive points occur at comparatively large fields and exposure-equivalent statistics. Their number depends on the chosen grid and does not represent a detection probability or population fraction.
The two effects therefore represent distinct limitations within the investigated model. The adopted systematic-error treatment reduces the sensitivity gain from increasing exposure, while continuum degeneracy reduces the distinguishability of the conversion signature from the source spectrum. Stronger magnetic fields can improve the conversion probability, but do not by themselves ensure competitive sensitivity. Independent information about the intrinsic continuum and improved characterization of instrumental and background uncertainties are therefore important for realizing the sensitivity suggested by the fixed-continuum, statistics-dominated scaling.
 
\begin{figure}[tbp]
\centering
\includegraphics[width=\textwidth]{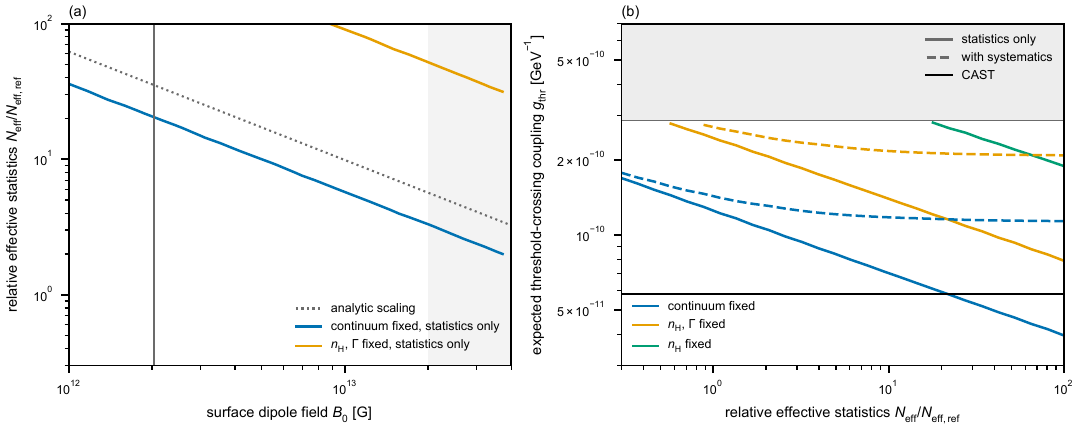}
\caption{Forward-folded sensitivity forecast for a J2229-like spectrum in the 1–6 keV band, assuming $f_O=1$ and $\theta=90^\circ$. (a) CAST-competitiveness boundaries in surface dipole field $B_0$ and exposure-equivalent multiplier $n$. The dotted line is the analytic fixed-continuum scaling anchored to the observed-data crossing; solid curves are interpolated statistics-only forecast boundaries. No grid point reaches the CAST bound under the adopted systematic-error treatment. The vertical line marks the J2229+6114 reference field, and the gray region marks the caution region above $2\times10^{13}\,\mathrm{G}$. (b) Threshold-crossing coupling at the nearest grid field to J2229+6114, $B_0=1.879\times10^{12}\,\mathrm{G}$. Solid and dashed curves represent statistics-only and systematic-error treatments; colors indicate the continuum treatment. Lines connect evaluated grid points. The horizontal black line is the CAST bound, and the gray region lies above the $10\%$ depletion cap. The fully profiled continuum does not reach the threshold in the displayed range.}
\label{fig:forecast}
\end{figure}

\begin{figure}[tbp]
\centering
\includegraphics[width=0.65\textwidth]{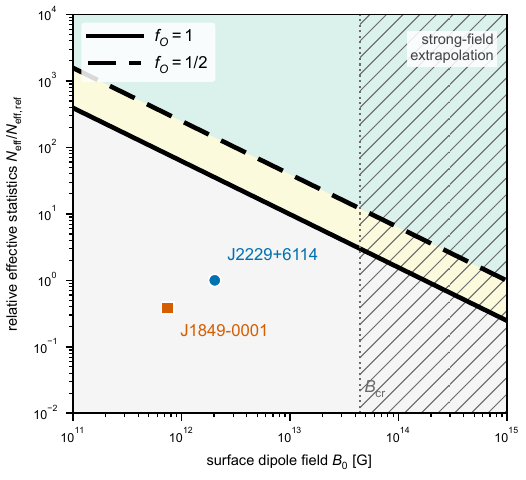}
\caption{Idealized CAST-competitiveness boundaries as a function of surface dipole field and relative effective statistics, using the fixed-continuum, statistics-dominated scaling of \eqref{eq:g95}. Solid and dashed lines correspond to $f_{O}=1$ and $f_{O}=1/2$, respectively. Above each line, the corresponding idealized sensitivity is stronger than the CAST comparison bound. The labeled J2229$+$6114 point defines the reference; the J1849$-$0001 point is placed at the effective-statistics multiplier that reproduces its conditional crossing and is not simply the exposure ratio. The hatched region above $B_{cr}$ represents a strong-field extrapolation and is not used for quantitative conclusions.}
\label{fig:castsens}
\end{figure}

\section{Conclusions}
\label{sec:Conclusion}

We searched the NICER spectra of PSR J2229$+$6114 and PSR J1849$-$0001 for X-ray photons lost to ALPs in a dipolar magnetosphere. QED vacuum birefringence suppresses coherent conversion near the stellar surface. In the low-mass regime studied here, the resulting depletion varies smoothly with energy, approximately as $E^{-4/5}$ in the baseline model. Over either fitted band, such a deficit can resemble a change in the slope and normalization of the absorbed power law. To measure how much of it the data can distinguish, we folded the same conversion template through each NICER response and repeated the coupling scans while changing only which continuum parameters were free. With all three parameters fixed, the maximum increase in $\chi^2$ relative to the near-zero coupling is 43.33 and 49.80 for the two sources, respectively. Fixing absorption alone gives no increase for either source. With all three parameters free, even the joint $\chi^2$ varies by only 0.27 over the entire common scan. Thus the data do not exclude a coupling in the tested perturbative range; treating the fitted continua as independently known would give a different answer on the very same photons.

The J2229-like forecast tests whether a stronger field or longer exposure could recover CAST sensitivity. With the continuum known and statistical errors alone, some combinations of field and exposure reach the benchmark. Retaining the adopted 1.5\% fractional uncertainty on the total counts per grouped bin changes the fitted exposure dependence of the fixed-continuum crossing from approximately $n^{-0.25}$ to $n^{-0.07}$ in the core field range. None of the 673 grid points then reaches CAST: the best $g_{\rm CAST}/g_{\rm thr}$ is 0.82 in the core range and 0.93 including the higher-field extension, where the weak-field approximation is less reliable. With the continuum free, no scan reaches the exclusion threshold under either uncertainty treatment, even at 100 times the reference exposure. The uncertainty of the 3C50 background prediction itself was not independently quantified. For smooth depletion in spectra of this kind, a projection of competitive sensitivity must therefore show that the residual curvature survives fitting the source continuum and remains resolvable at the achieved spectral precision.

\section*{Acknowledgments}
We thank the Institute of Physics, Academia Sinica, and the Department of Physics, National Cheng Kung University, for their support. We also acknowledge funding from the National Science and Technology Council (NSTC), Taiwan.
\clearpage

\appendix
\section{Summary of Observations}
\label{app:data_table}

\begin{table}[h]
\centering
\begin{tabular}{ccccc}
\hline \hline
 Pulsar Name & Surface Magnetic Field (G) & ObsID & Exposure Time (s) \\ \hline
 \multirow{10}{*}{J1849-0001} & \multirow{10}{*}{$7.5\times10^{11}$} & 
 3536040701 & 11306 \\
 & & 2579040401 & 10321 \\
 & & 3536041102 & 9740 \\
 & & 3536040101 & 9451 \\
 & & 5505050703 & 8777 \\
 & & 5505050501 & 8383 \\
 & & 3536040806 & 8315 \\
 & & 5505051301 & 8261 \\
 & & 5505051101 & 8258 \\
 & & 5505051003 & 7798 \\ \hline
 \multirow{10}{*}{J2229+6114} & \multirow{10}{*}{$2.03\times10^{12}$} 
 & 2579050301 & 30362 \\
 & & 2579050914 & 24539 \\
 & & 2579050401 & 22299 \\
 & & 3536051101 & 19955 \\
 & & 2579050202 & 19110 \\
 & & 2579050302 & 19092 \\
 & & 2579051210 & 18607 \\
 & & 2579051101 & 18530 \\
 & & 2579050101 & 18176 \\
 & & 1033370139 & 17104 \\
\hline \hline
\end{tabular}
\caption{\label{tab:data} Summary of the NICER observations used in this analysis. The surface  fields ($B_0$) are given in Table~\ref{tab:obs}; ObsID identify cleaned segments used for fitting.}
\label{app:obs}
\end{table}

\section{Numerical validation of the conversion template}
\label{app:B}
We checked the template $\hat{P}$against a direct integration of the photon–ALP evolution equations along a radial ray at $\theta=90^{\circ}$,

\begin{equation}
i\frac{d}{dr}\begin{pmatrix} A_{\gamma} \\ A_{a}\end{pmatrix} = \begin{pmatrix} \Delta_{pl}+\Delta_{QED} & \Delta_{a\gamma} \\ \Delta_{a\gamma} & \Delta_{a} \end{pmatrix}\begin{pmatrix} A_{\gamma} \\ A_{a} \end{pmatrix}
\end{equation}

with
\[
\begin{aligned}
\Delta_{a\gamma} &= \frac{g_{a\gamma\gamma}B_{\perp}}{2}, &
\Delta_{a} &= -\frac{m_{a}^{2}}{2\omega_{loc}},\\
\Delta_{pl} &= -\frac{\omega_{pl}^{2}}{2\omega_{loc}}, &
\Delta_{QED} &= \frac{7\alpha_{em}}{90\pi}\left(\frac{B_{\perp}}{B_{cr}}\right)^{2}\omega_{loc}.
\end{aligned}
\]
The field is a dipole, $\omega_{pl}$ follows the Goldreich–Julian density scaled by the pair multiplicity $\kappa$, and the local energy is $\omega_{loc}=\omega/\alpha(r)$ with $\alpha(r)=(1-r_{s}/r)^{1/2}$.The factor $1/\alpha(r)$ converts coordinate to proper radial length. Integration starts at $R_{NS}$, stops at the light cylinder, and uses an adaptive fourth-order Runge–Kutta scheme with step doubling.
Explicit Runge–Kutta schemes do not conserve the norm of an oscillatory system, so we tightened the tolerances until $|A_{\gamma}|^{2}+|A_{a}|^{2}$ was conserved to $10^{-4}$ at the highest energy. A further factor of ten changed $P_{\gamma \to a}$ by less than $10^{-4}$. Independently, a first-order evaluation of the same equations agrees with the integration to $2\times10^{-4}$ for every mass and multiplicity considered. At these couplings $(P_{\gamma\to a}\sim 10^{-6})$ the first-order treatment is exact for practical purposes.
The figure compares the integration with the template. At $\kappa=1$ and $m_{a}=10^{-6}$eV, the integrated probability exceeds the template quadrature by an energy-independent factor of 1.163-1.166 for both pulsars. At the sampled energies within the fitted bands, it exceeds the analytic form $AE^{-0.8}$ by approximately $10\%$-$13\%$; the residual $\pm 2\%$ trend across the band comes from the truncation at the light cylinder. The offset arises because the template evaluates the gravitational redshift at the stellar surface, $\alpha(R_{NS})=0.83$, whereas conversion occurs at $r\sim20R_{NS}$, where $\alpha\simeq0.99$; the template therefore overestimates the QED phase. A constant offset is degenerate with $g_{a\gamma\gamma}^{2}$, and since the template lies below the integration it is conservative: with the integrated normalization, the quoted fixed-continuum couplings would decrease by about $5\%$ at small depletion, and by less near the cap (see below). For $\kappa=10^{3}$ the comparison is unchanged within $2\%$. For $\kappa=10^{4}$ the low-energy enhancement is larger than the template predicts (ratio up to 1.33 at 1 keV for J2229$+$6114), because the surface-redshift approximation also suppresses the plasma phase. The enhancement increases the conversion probability, so it does not weaken the $\kappa=1$ results.
Close to the depletion cap the first-order form $1-g^{2}\hat{P}$ is no longer exact. At $g_{cap}$ the integrated $P_{\gamma\to a}$ lies $8.2\%$ below the $g^{2}$ extrapolation of the weak-coupling result at the bottom of each band, $1.3$-$4\%$ below it at higher energies, and $10\%$ below it at $\kappa=10^{4}$. This partly cancels the normalization offset: at $g_{cap}$ the true depletion at the bottom of the band is $10.4\%$ for both pulsars, compared with the template's $10\%$, and exceeds the template at every band energy for $m_{a}=10^{-6}$eV. At $m_{a}=10^{-5}$ eV it is $9.6\%$ (J2229$+$6114, 1 keV) and $10.1\%$ (J1849$-$0001, 2 keV), so at the cap the template is accurate to $4\%$ but not strictly conservative. At the all-fixed crossings the depletion is $2$-$3\%$ and the nonlinear correction is about $2\%$.

To test the mass dependence, we repeated the integration for $10^{-7}\leq m_{a}\leq10^{-4}$ eV at 6 (J2229$+$6114) and 4 (J1849$−$0001) energies across each fit band, with $\kappa=1$. Relative to $m_{a}=10^{-6}$ eV, the conversion probability changes by less than $0.6\%$ for $m_{a}\leq3\times10^{-6}$ eV. At $m_{a}=10^{-5}$ eV it decreases by at most $7.6\%$ (J2229$+$6114, 1 keV) and $2.8\%$ (J1849$−$0001, 2 keV); after a free normalization, the residual change in spectral shape is $3.9\%$ and $1.1\%$. At $m_{a}=10^{-5}$ eV the integrated probability exceeds $AE^{-0.8}$ at the sampled energies in both bands, by at least $4.6\%$. Above $m_{a}\simeq1.5\times10^{-5}$ eV the mass phase suppresses low-energy conversion rapidly. At $2\times10^{-5}$ eV the probability at 1 keV for J2229+6114 is $32\%$ below the $10^{-6}$eV value and $23\%$ below the template, and the template no longer applies. The mass dependence is weaker at smaller field angles (at $\theta=8^{\circ}$, at most $3.9\%$ at $10^{-5}$ eV), so $\theta=90^{\circ}$ is the limiting case. We therefore restrict all quantitative statements to $m_{a}\leq10^{-5}$ eV.

\begin{figure}[tbp]
\centering
\includegraphics[width=\textwidth]{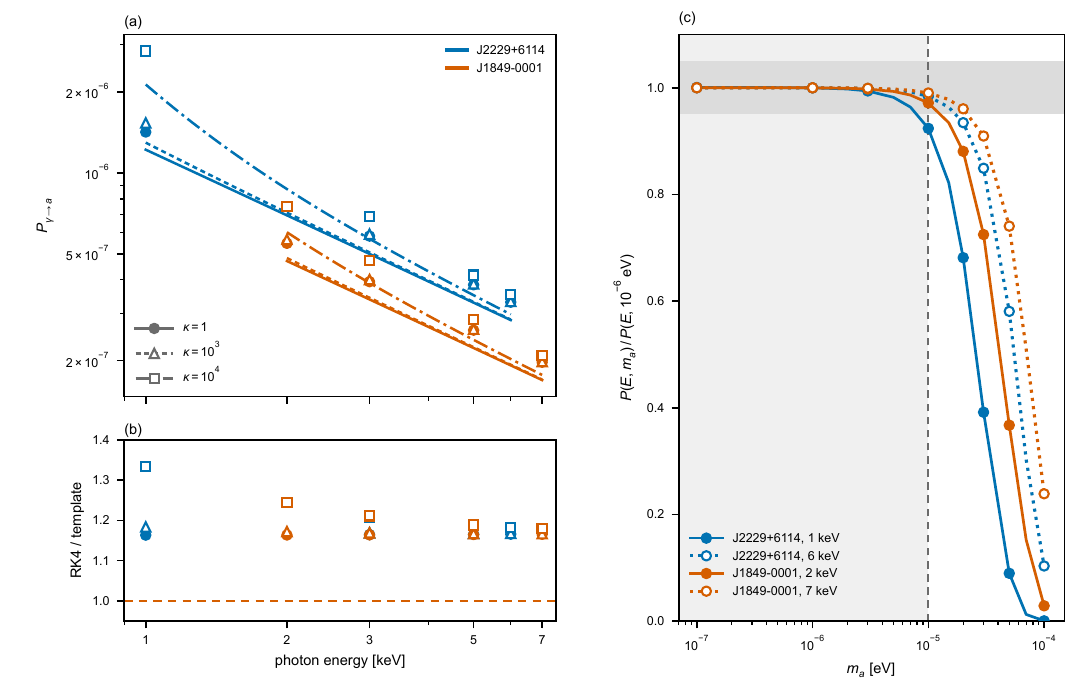}
\caption{Validation of the conversion template against numerical propagation for $\theta=90^{\circ}$ and $g_{a\gamma\gamma}=10^{-12}\mathrm{GeV}^{-1}$. (a) Conversion probability at $m_{a}=10^{-6}$eV; lines show the template quadrature and markers show the RK4 result for $\kappa=1$, $10^{3}$ and $10^{4}$. (b) Ratio of the RK4 result to the template quadrature. (c) Mass dependence at $\kappa=1$, normalized to the probability at $m_{a}=10^{-6}$eV; filled and open markers correspond to the lower and upper edges of each fitted band, respectively. Lines show first-order calculations and markers show RK4 results. The vertical shading indicates the displayed low-mass reference region up to $10^{-5}$eV, and the horizontal band marks $\pm5\%$. The J2229$+$6114 1 keV result at $m_{a}=10^{-5}$eV lies below that band.}
\label{fig:rk4}
\end{figure}

\section{Auxiliary conditional-reach plots}
\label{app:C}
\begin{figure}[!htbp]
\centering
  \includegraphics[width=\textwidth]{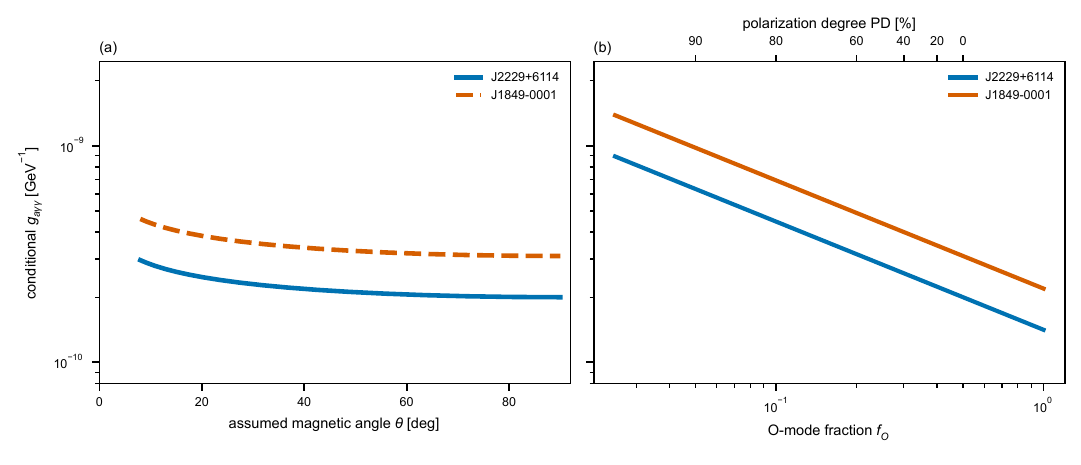}
\caption{Analytic rescaling of the fixed-continuum threshold-crossing coupling under amplitude-only parameter variations. (a) Dependence on the assumed magnetic-field angle $\theta$, with $f_O=1/2$. (b) Dependence on the constant O-mode fraction $f_O$ at $\theta=90^{\circ}$; the upper axis indicates the polarization degree on the X-mode-dominated branch $f_O\leq1/2$.}
\label{fig:sys}
\end{figure}

The fully fixed-continuum reference of Section~\ref{sec:profile-result} can be rescaled analytically for parameters that modify only the amplitude of the low-mass template. Figure~\ref{fig:sys} illustrates two such scalings, for the assumed magnetic-field angle and the constant O-mode fraction.

\bibliographystyle{unsrt}
\bibliography{reference}

\end{document}